\documentclass[reprint,showpacs,amsmath,amssymb,twocolumn,letterpaper,aps,prb]{revtex4-1}

\usepackage{graphicx}% Include figure files
\usepackage{times}
\usepackage{dcolumn}
\usepackage{eqnarray}
\usepackage{varwidth}   

\begin{document}

\title{Investigation of the complex magnetism in the breathing pyrochlore LiIn(Cr$_{1-x}$Rh$_x$)$_4$O$_8$}
%\thanks{The research performed in this study was supported by the Ministry of Science and Technology (MOST) under grant no. 2016YFA0300503.}

%\author{D. Wang}
%\affiliation{State Key Laboratory of Surface Physics, Department of Physics, Fudan University, Shanghai 200433, China}
%\author{C. Tan}
%\affiliation{State Key Laboratory of Surface Physics, Department of Physics, Fudan University, Shanghai 200433, China}
%\author{K. Huang}
%\altaffiliation[Corresponding Authors: ]{khuang@fudan.edu.cn (Phone: 186-2622-4375), leishu@fudan.edu.cn (Phone: 5163-0252)}
%\affiliation{State Key Laboratory of Surface Physics, Department of Physics, Fudan University, Shanghai 200433, China}
%\author{L. Shu}
%\altaffiliation[Corresponding Authors: ]{khuang@fudan.edu.cn (Phone: 186-2622-4375), leishu@fudan.edu.cn (Phone: 5163-0252)}
%\affiliation{State Key Laboratory of Surface Physics, Department of Physics, Fudan University, Shanghai 200433, China}
%\affiliation{Collaborative Innovation Center of Advanced Microstructures, Nanjing 210093, China}

\author{D. Wang$^1$}
\author{C. Tan$^1$}
\author{K. Huang$^1$}\email[Corresponding Author: ]{leishu@fudan.edu.cn, khuang@fudan.edu.cn}
\author{L. Shu$^1$$^,$$^2$}
\email[Corresponding Author:]{khuang@fudan.edu.cn, leishu@fudan.edu.cn,}

\affiliation{$^1$State Key Laboratory of Surface Physics, Department of Physics, Fudan University, Shanghai 200433, China}
\affiliation{$^2$Collaborative Innovation Center of Advanced Microstructures, Nanjing 210093, China}

\begin{abstract}
We have performed a detailed investigation of the new `breathing' pyrochlore compound LiInCr$_4$O$_8$ through Rh substitution with measurements of magnetic susceptibility, specific heat, and x-ray powder diffraction. The antiferromagnetic phase of LiInCr$_4$O$_8$ is found to be slowly suppressed with increasing Rh, up to the critical concentration of $x$ = 0.1 where the antiferromagnetic phase is still observed with the peak in specific heat $T_p$ = 12.5 K, slightly lower than $T_p$ = 14.3 K for the $x$ = 0 compound. From the measurements of magnetization we also uncover evidence that substitution increases the amount of frustration. Comparisons are made with the LiGa$_y$In$_{1-y}$Cr$_4$O$_8$ system as well as other frustrated pyrochlore-related materials and find comparable amounts of frustration. The results of this work shows that engineered breathing pyrochlores presents an important method to further understand the complex magnetism in frustrated systems.
\end{abstract}


\maketitle

%\section{Introduction}
Magnetic frustration has recently attracted renewed interest as novel and exotic new phases can arise from the frustrated magnetic interactions.\cite{GARDNER10} The pyrochlore class of materials is an ideal family to investigate as its crystal structure promotes magnetic frustration. Forming conventionally in the $A_2 B_2$O$_7$ composition, the $A$ and $B$ atoms form corner-sharing tetrahedras. If the nearest neighbor exchange interaction for the $A$ and $B$ atoms are antiferromagnetic then there are no configurations for which the magnetic moments can simultaneously satisfy all nearest neighbor interactions, geometrically promoting magnetic frustration. It is no surprise then that the pyrochlores have displayed a wide range of interesting phenomena such as spin ice and spin glasses,\cite{BRAMWELL01, RAJU92} metal-insulator transitions,\cite{MANDRUS01} potential topological insulators,\cite{YANG10} and superconductivity.\cite{YONEZAWA04, YONEZAWA04a, YONEZAWA04b}

The pyrochlore family has been known for a long time, first discovered in the 1930's.\cite{GARDNER10, GAERTNER30} However, it was only after the discovery of spin-glass like properties in Y$_2$Mo$_2$O$_7$ did the family's unique potential for novel magnetic properties become realized.\cite{GARDNER10, GREEDAN86} Recently there have been several new exotic entries into the pyrochlore family. Discovered in 2015, the $RE_3$Sb$_3$Zn$_2$O$_{14}$ branch was the first member to display the 2D kagome lattice,\cite{SANDERS16} which has a high potential for exhibiting the exotic spin-liquid state. Another recent discovery was the `breathing' pyrochlores and will be the focus of this work.

The breathing pyrochlores were discovered in 2013 and the first materials formed in the chemical composition Li$M$Cr$_4$O$_8$, $M$ = In or Ga,\cite{OKAMOTO13} a variant of the conventional pyrochlore structure. The Li and $M$ atoms alternate in series and due to the large differences in size produce a lattice that periodically expands and contracts, the origin of the breathing term. Investigations on Li$M$Cr$_4$O$_8$ ($M$ = In or Ga) reveal unusual magnetic and electronic properties where both compounds show a magnetic phase transition tied to structural distortions at 13.8 K and 15.9 K for $M$ = Ga and In, respectively.\cite{OKAMOTO13} Nuclear magnetic resonance reveals a more complicated phase diagram with spin-gap, structural, and long range magnetic order in LiInCr$_4$O$_8$ while LiGaCr$_4$O$_8$ shows no spin-gap yet displays a potential tri-critical point.\cite{TANAKA14} A more recent investigation using multiple spin resonance techniques (electron, nuclear, and muon) show that LiGaCr$_4$O$_8$ has a magnetostructural phase transition at 15.2 K followed by the long-range magnetic order at 12.9 K while LiInCr$_4$O$_8$ crosses over from a correlated paramagnet with a weak magnetostructural transition at 17.6 K and long range magnetic order at 13.7 K.\cite{LEE16} Furthermore, a spin-glass like phase develops in LiGa$_{y}$In$_{1-y}$Cr$_4$O$_8$ at moderate substitutions after the antiferromagnetism of either end member is suppressed as well as a `pseudo' spin-gap behavior observed in the near the critical concentration of $y$ = 0.1.\cite{OKAMOTO15}

%Being another approach to exotic states, bond alternation is mainly studied in one-dimensional antiferromagnetic systems. In simplified one-dimensional systems, there are two bonds alternating along the chain, one with strong antiferromagnetic interactions $J$ and the other with weak ones $J$'( $J$'/$J$$<$1). A uniform spin-1/2 chain shows a gapless Tomonaga-Luttinger liquid state , whereas a spin-gapped state with a spin singlet pair on the J bond occurs under the circumstance of a finite alternation. At the same time, a gapped Haldane state appears for zero or small alternations in spin-1 chains while a dimer singlet state is stabilized at large alternations. With completely different physical origins of these spin gaps, there exists a gapless state at a quantum critical point $J$'/$J$=0.6 between the two phases.

The pseudo spin gap behavior is observed with small Ga substitution from the LiInCr$_4$O$_8$ parent when the antiferromagnetic phase is fully suppressed. Therefore, we have performed chemical substitution of LiInCr$_4$O$_8$ with Rh substituted on the Cr site as LiInRh$_4$O$_8$ is reported to be non-magnetic.\cite{OKAMOTO15} From our results we find that the peak $T_p$ in specific heat due to the antiferromagnetic phase is slowly suppressed with increasing Rh up to $x$ = 0.1, an unexpected result as Rh substitution should significantly alter the electronic configuration. Furthermore we find evidence that the frustration is enhanced with chemical substitution and comparisons are made to other frustrated systems. We find that the frustration generated from the engineered breathing pyrochlores are comparable to conventional geometrically frustrated systems, such as ZnCr$_2$O$_4$,\cite{OKAMOTO13, MARTINHO01} providing an important route for further understanding the complex magnetism in frustrated systems.

%We find that the frustrated generated from the breathing lattice are comparable to other more conventional materials such as ZnCr$_2$O$_4$,\cite{OKAMOTO13, MARTINHO01} showing that engineering the `breathing' lattices is a viable route towards highly frustrated materials.

%\section{Experimental Details}

Polycrystalline samples of LiIn(Cr$_{1-x}$Rh$_x$)$_4$O$_8$ with $x$ = 0, 0.025, 0.5, 0.075, and 0.1, were synthesized by solid state reaction in a conventional Lindberg box furnace. The starting constituent materials are Li$_2$CO$_3$, In$_2$O$_3$, and Cr$_2$O$_3$/Rh$_2$O$_3$ which were dried over night in at 120 C and then weighed out to the molar ratio of 1:1:4. The starting materials were then mechanically mixed, pressed into pellets, and sintered for 48 hours at 1100 C$^\circ$. After heating the samples were crushed into powder, re-pressed into pellets, and sintered up to four more times to ensure homogeneity. Powder x-ray diffraction measurements were performed on all samples using a Bruker D8 Discover x-ray diffractometer with a Cu K$_{\alpha}$ source. Magnetization measurements were performed using a Quantum Design Vibrating magnetometer from 300 K down to 2 K in applied magnetic fields up to 5 Tesla. Specific heat measurements were performed in a Quantum Design Physical Properties Measurement System Dynacool which employs a standard thermal relaxation technique.

%\section{Experimental Results}

\begin{figure}[ht]
 \begin{center}
 \includegraphics[width=0.4\textwidth]{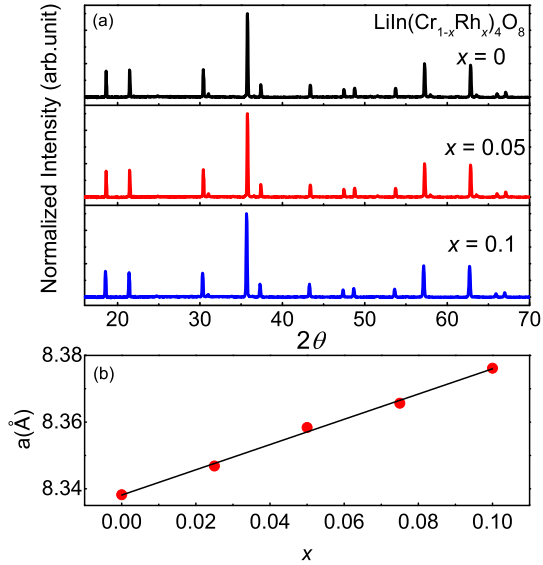}
 \caption{ (Colour online) (a) Powder x-ray diffraction pattern for representative concentrations of LiIn(Cr$_{1-x}$Rh$_x$)$_4$O$_8 $. The black, red, and blue data sets correspond to $x$ = 0, 0.05, and 0.1 respectively. The measured intensity have been normalized to the highest peak at 2$\theta$ = 36$^{\circ}$ and the data sets have been offset for clarity. (b) The lattice parameter $a$, vs. Rh concentration $x$. For all measured concentrations the lattice parameter increases linearly with increasing $x$, starting from 8.338 $\text{\AA}$ for $x$ = 0 up to 8.373 $\text{\AA}$ at $x$ = 0.1 and was well described by the expression $a$ = 8.3381(8) + 0.379(13)$x$, represented by the black solid line. }
 \label{fig:xray}
 \end{center}
\end{figure}

The x-ray diffraction patterns for representative concentrations are displayed in Fig.~\ref{fig:xray} with the data sets normalized to the highest peak intensity at 2$\theta$ = 36$^{\circ}$.
Rietveld refinements were performed on the powder XRD patterns for each sample using GSAS\cite{LARSON04} and EXPGUI\cite{TOBY01}. All of the x-ray diffraction data sets are consistent with a cubic $F\bar{4}3m$ crystal structure and the peak positions well fit with the theoretical peak positions. The lattice parameter, $a$, increases linearly with increasing Rh concentration and is displayed in Fig.~\ref{fig:xray}(c). Furthermore, $a$ follows the relation $a$ = 8.3381(8) + 0.379(13) $x$ and is represented by the solid black line.

\begin{figure}[ht]
 \begin{center}
 \includegraphics[width=0.5\textwidth]{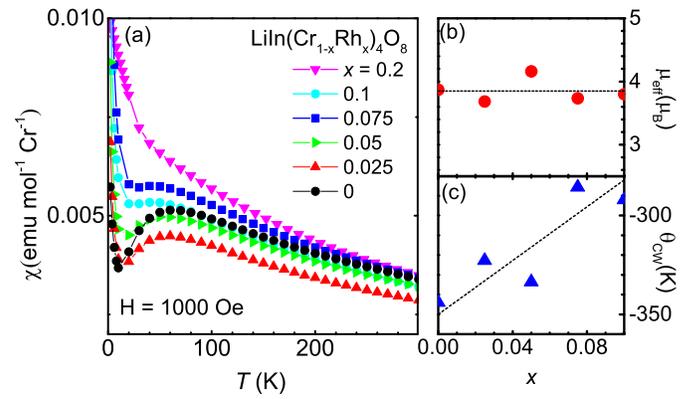}
 \caption{ (Colour online) (a) Temperature dependence of $\chi$, for different concentrations of Rh from $x$ = 0 up to $x$ = 0.2 with an applied magnetic field of 1000 Oe. Curie-Weiss fits were performed on temperatures above 100 K. The results of $\mu_{eff}$ and $\theta_{CW}$ are displayed in panels (b) and (c), respectively. $\mu_{eff}$ appears to be independent of $x$, staying near 3.85 $\mu_{B}$, represented by the black dashed line in panel (b). $\theta_{CW}$ on the other hand shows a positive dependence with $x$, increasing linearly with increasing $x$ for all samples measured. The black dashed line in panel (c) is a guide to the eye.}
 \label{fig:mag}
 \end{center}
\end{figure}

Illustrated in Fig.~\ref{fig:mag}(a) is the magnetic susceptibility, $\chi$ vs. $T$ for all measured Rh concentrations in an applied magnetic field of 1000 Oe. $\chi$ is displayed as per Cr atom as Rh is expected to be non-magnetic.
For concentrations up to $x$ = 0.1 there is a broad peak at roughly 50 K that decreases in magnitude with increasing $x$, which becomes more broad and slightly suppressed with increasing Rh substitution. Above 100 K the magnetic susceptibility displays Curie-Weiss behavior $\chi$ = $C$/($T$-$\theta_{CW}$). Displayed in Fig.~\ref{fig:mag}(b) and (c) are the determined values for $\mu_{eff}$ ($\mu_{eff}$ $\propto$ 2.83 $\sqrt{C}$) and $\theta_{CW}$, respectively. The value of $\mu_{eff}$/Cr stays constant near 3.85 $\mu_{B}$, close to the Hund's rule value of 3.87 $\mu_{B}$, evidence that there is no spin-orbit coupling in this system. $\theta_{CW}$ increases linearly with increasing Rh concentration, approaching lower negative values from $-340$ K for $x$ = 0 up to -290 K for $x$ = 0.1, evidence the system is becoming less antiferromagnetic with increasing Rh. At low temperatures $\chi$ appears to diverge, which has previously been attributed to orphan spins/magnetic impurities of 0.2\%.\cite{OKAMOTO15}

\begin{figure}[ht]
 \begin{center}
 \includegraphics[width=0.4\textwidth]{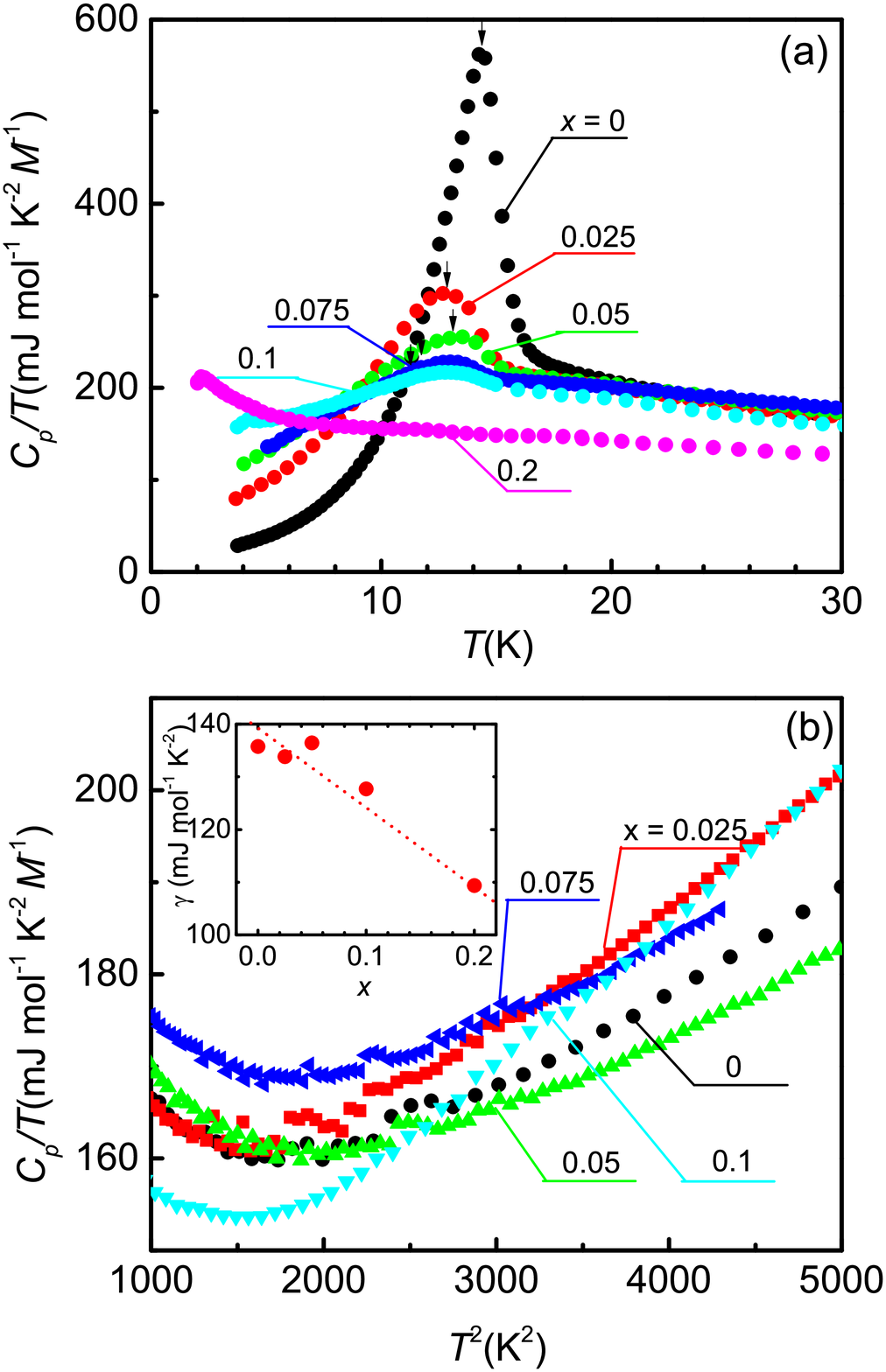}
 \caption{ (Colour online) (a) Specific heat data displayed as $C_p/T$ vs. $T$ for samples with Rh concentrations up to $x$ = 0.2. A pronounced peak is observed at 14 K for the $x$ = 0 sample and systematically decreases in amplitude with increasing Rh up to $x$ = 0.1 where the peak becomes a broad feature and by $x$ = 0.2 no broad feature can be observed. (b) The specific heat data displayed as $C_p$/$T$ vs. $T^2$ to highlight the linear behavior above $\sim$3000 K$^2$ (the $x$ = 0.2 data set has been omitted for clarity). The inset displays $\gamma$ as a function of $x$ with the dashed red line serving as a guide to the eye.}
 \label{fig:heat}
 \end{center}
\end{figure}

Fig.~\ref{fig:heat}(a) displays the specific heat data as $C_p/T$ vs. $T$ for concentrations of $x$ up to 0.2. The data is displayed per $M$ atom ($M$ = Cr or Rh) as both elements would contribute. For the $x$ = 0 sample a sharp peak is observed at $T_p$ = 14.3 K (indicated in the graph by the black arrow). Initial Rh substitution rapidly suppresses $T_p$ but has a significantly diminished effect with further substitution as $T_p$ drops to 12.7 K for x = 0.025 but stays almost constant for higher $x$ as $T_p$ = 12.5 K for $x$ = 0.1. Interestingly the suppression of $T_p$ is noticeably slower than that observed in the LiGa$_{y}$In$_{1-y}$Cr$_4$O$_8$ system, which shows the complete suppression of $T_p$ at 6$\%$ Ga substitution.\cite{OKAMOTO15} It should be noted that previous investigations on LiInCr$_4$O$_8$ observed two features the specific heat data, a sharp peak associated with a structural phase transition at $T_p$ = 15.9 K and a shoulder at $T_S$ = 14 K, which was associated with the antiferromagnetic transition. However, the same study also found that the doped samples only displayed the peak which became associated with the antiferromagnetic ordering, even in concentrations as small as 2.5$\%$.\cite{OKAMOTO15} Combined with the high sensitivity of $T_p$ to initial substitution (the 2.5$\%$ Ga substituted sample $T_p$ = 12.9 while the $x$ = 0 sample in this study displays $T_p$ = 14.3 K, much closer to $T_p$ = 15.9 K of the previously reported $x$ = 0) suggests that the absence of a shoulder in the $x$ = 0 sample is most likely due to trace amounts of impurities. Additionally a small peak is observed at 4.2 K and 2.2 K for $x$ = 0.1 and 0.2 respectively, which appears to be a separate feature from $T_p$ as it is still clearly observed for $x$ = 0.1 at 12.5 K.

%Increasing Rh concentration suppresses $T_p$ to lower temperatures and becomes less pronounced with $T_p$ = 12.5 K for $x$ = 0.1 and by $x$ = 0.2 (not shown) the peak is no longer observed down to 2 K.

Displayed in Fig.~\ref{fig:heat}(b) is the specific heat data plotted as $C_p/T$ vs $T^2$.
For all measured samples the data appear linear above roughly 3000 K$^2$ ($\sim$50 K) and are well described by $C$ = $\gamma T$ + $\beta T^3$, where the first and second terms correspond to the electronic and phonon contributions, respectively. This can be seen in Fig.~\ref{fig:heat}(b) which displays the same data as Fig.~\ref{fig:heat}(a) plotted as $C_p/T$ vs $T^2$, except for $x$ = 0.2 which was omitted for clarity.  The results for $\gamma$ are displayed in the inset of Fig.~\ref{fig:heat}(b) and appears to decrease linearly with increasing Rh concentration, starting from $\sim$135 mJ / mol K$^2$ for low concentrations of Rh and decreasing down to 109 mJ / mol K$^2$ for $x$ = 0.2.

%\section{Discussion}

\begin{figure}[ht]
 \begin{center}
 \includegraphics[width=0.4\textwidth]{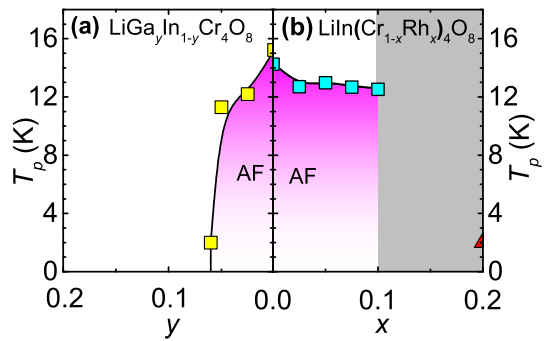}
 \caption{$T_p$ plotted as a function of chemical substitution for both series LiGa$_{y}$In$_{1-y}$Cr$_4$O$_8$ from Ref.\onlinecite{OKAMOTO15} in panel (a) and for LiIn(Cr$_{1-x}$Rh$_x$)$_4$O$_8$ displayed in panel (b). The red triangle in panel (b) represents the specific measurements on the $x$ = 0.2 sample which showed no magnetic order down to 2 K. The grey region in panel (b) represents the region in the phase diagram that has not yet been investigated. The regions labelled AF refers to the antiferromagnetic long-range ordered phase.}
 \label{fig:phase}
 \end{center}
\end{figure}

Displayed in Fig.~\ref{fig:phase} is $T_p$, the peak associated with the antiferromagnteic transition, vs. chemical substitution for LiGa$_y$In$_{1-y}$Cr$_4$O$_8$ taken from reported literature,\cite{OKAMOTO15} in panel (a) and for LiIn(Cr$_{1-x}$Rh$_x$)$_4$O$_8$ in panel (b). Immediately it becomes clear that the different chemical substitutions produce different responses in $T_p$. Rh substitution seems to have very little effect on $T_p$, only decreasing from 14.3 K for $x$ = 0 down to 12.5 K for $x$ = 0.1 while LiGa$_y$In$_{1-y}$Cr$_4$O$_8$ displayed a more rapid suppression of $T_p$ where $y$ = 0.06 completely suppressed $T_p$. The grey area in Fig.~\ref{fig:phase}(b) represents the region of $x$ not investigated in this study which contains $x_{cr}$, the critical concentration where $T_p$ is fully suppressed.

One possible explanation for the observed change in $T_p$ with chemical substitution is the change in electronic configuration of the substituted elements. Ga-In is an iso-valent substitution where both elements have very similar electronic configurations, Ga displays $3d^{10}4s^24p^1$ while $4d^{10}5s^24p^1$ for In, with no change in the amount of valence electrons. This is contrasted with Cr and Rh which significantly changes the electronic configuration where Cr exhibits $3d^54s^1$ and Rh is $4d^85s^1$, which in a simplistic view means Rh substitution adds 3 electrons. However, the change in electronic configuration is unlikely to explain the observed changes in $T_p$, as the iso-valent substitution of Ga-In results in a more rapid drop of $T_p$ while adding electrons through Rh substitution of Cr results in almost no change in $T_p$, staying near 12.5 K.

\begin{table*}[ht]
\caption{Listed are the Curie-Weiss temperatures and characteristic temperature $T^*$ (i.e. $T_p$, spin-glass temperature, spin-singlet crossover) for several relevant pyrochlores.}\label{tab:fratio}
%\begin{tabular}{l c c c c}
\begin{tabular}{m{4.5cm} m{2cm} m{2cm} m{1.5cm} c}
\hline
 & $\theta_{CW}$ (K) & $T^{*}$ (K) & $f$ & Reference \\
\hline\hline
LiInCr$_4$O$_8$ & -344 & 15 & 22.6 & This work \\
LiIn(Cr$_{0.9}$Rh$_{0.1}$)$_4$O$_8$ & -292 & 14 & 20.8 & This work \\
LiGaCr$_4$O$_8$ & -656 & 13.8 & 47 & [\onlinecite{OKAMOTO13}] \\
Ba$_3$Yb$_2$Zn$_5$O$_{11}$ & -128 & 4 & 32 & [\onlinecite{KIMURA14}] \\
Ba$_2$Sn$_2$Ga$_3$ZnCr$_7$O$_{22}$ & -315 & 1.5 & 200 & [\onlinecite{HAGEMANN01}] \\
ZnCr$_2$O$_4$ & -388 & 12 & 25 & [\onlinecite{OKAMOTO13, MARTINHO01}] \\
\hline
\end{tabular}
\end{table*}

Another potential explanation is the effect of chemical pressure to describe the change in magnetic properties. The variation in the unit cell can be used to estimate an equivalent amount of chemical pressure, $P_{ch}$, according to the isothermal compressibility $\kappa_T$ (or bulk modulous $B_0$ = 1/$\kappa_T$), as has been used in other materials such as URu$_2$Si$_2$ with Fe substitution.\cite{KANCHANAVATEE11} While $\kappa_T$ for the breathing pyrochlores are unknown, the compressibility of the related spinel oxides are known and exhibit an almost universal value for $B_0$ (and therefore $\kappa_T$ as $k_T$ = 1/$B_0$),\cite{RECIO01} including that of ZnCr$_2$O$_4$ with $B_0$ = 173 – 210 GPa,\cite{ZHANG11} which we use as an estimate for the compressibility of the breathing pyrochlores. Comparing the chemical pressure from the critical concentrations of 10$\%$ Rh substitution and 6$\%$ Ga substitution may explain the different responses of $T_p$ to the different chemical substitutions. From this analysis we find a negative pressure for Rh substitution with $P_{ch}$ ranging from -0.07 MPa to -0.08 MPa for 10$\%$ Rh substitution. For Ga substitution the change in lattice results in a positive chemical pressure with $P_{ch}$ = 0.03-0.04 MPa for 6$\%$ Ga substitution. Recall that LiInCr$_4$O$_8$ is already near the limit of an isolated tetrahedral with $J$’/$J$ = 0.1 (LiGaCr$_4$O$_8$ exhibits $J$’/$J$ = 0.6), where $J$’ and $J$ are the nearest-neighbor magnetic interactions of the large and small tetrahedras formed by the Cr atoms.\cite{OKAMOTO13, OKAMOTO15} Negative pressure from Rh substitution would reduce $J$’ but as $J$’/$J$ = 0.1 and is already close to the limit of 0, additional negative pressure would have diminished effects such as a smaller change in $T_p$. On the other hand the upper limit of $J$’/$J$ = 1 is far off and therefore would not reduce the effects of positive chemical pressure from Ga substitution.

To better understand relationship between the frustration and magnetic order in the breathing pyrochlores, it is important to characterize the amount of frustration. The previous investigation into LiGa$_{y}$In$_{1-y}$Cr$_4$O$_8$ characterized the frustration by the breathing factor $B_f$ = $J'/J$, with $B_f$ = 0.6 for LiGaCr$_4$O$_8$ and a much smaller $B_f$ = 0.1 for LiInCr$_4$O$_8$.\cite{OKAMOTO13, OKAMOTO15} However, as this investigation directly alters the Cr occupying site with Rh substitution, it complicates the determination of $B_f$. Therefore in this study the frustration was instead characterized by the following equation  $f$ = $-\theta_{CW}$/$T^*$,\cite{HAGEMANN01} where $\theta_{CW}$ is determined from Curie-Weiss fits to the magnetic susceptibility and $T^{*}$ is the magnetic transition temperature, such as the N\'{e}el temperature for an antiferromagnet or spin-glass temperature. In this system, $T^*$ = $T_p$, the peak determined from specific heat. From this analysis we find that $f$ = 22.6 for $x$ = 0 and slowly decreases with increasing Rh reaching 20.8 for $x$ = 0.1 (there was no clear feature to determine $T_{p}$ for $x$ = 0.2). Performing the same analysis of $f$ on LiGa$_{y}$In$_{1-y}$Cr$_4$O$_8$ we find that Ga substitution increases $f$ where $f$ = 35 for $y$ = 0.05 and approaches $f$ = 47 for LiGaCr$_4$O$_8$, consistent with the increase in $B_f$ observed in the previous investigation of LiGa$_{y}$In$_{1-y}$Cr$_4$O$_8$.\cite{OKAMOTO13, OKAMOTO15}

Both LiInCr$_4$O$_8$ and LiGaCr$_4$O$_8$ are engineered systems in that the frustration was introduced through the `breathing' lattice, but importantly the amount of frustration appears to be comparable to traditional frustrated materials. For example, the well known frustrated system ZnCr$_2$O$_4$ displays an $f$ = 25, similar to that of LiInCr$_4$O$_8$ and actually is less frustrated than LiGaCr$_4$O$_8$ with $f$ = 47.\cite{OKAMOTO13, MARTINHO01} And using the singet-triplet crossover temperature $T^*$ = 4 K, the other known breathing pyrochlore Ba$_3$Yb$_2$Zn$_5$O$_{11}$ displays an $f$ = 32,\cite{KIMURA14} comparable to the frustration in Li$MT_4$O$_8$. However, it should be noted that there are other materials with much larger $f$, such as the 2-D spinel based Ba$_2$Sn$_2$Ga$_3$ZnCr$_7$O$_{22}$ which exhibits a much higher ratio of $f$ = 200.\cite{HAGEMANN01}

%It is noteworthy chemical substitution can have such differing effects on the amount of frustration. The isovalent substitution of LiGa$_{y}$In$_{1-y}$Cr$_4$O$_8$ results in a significant enhancement of frustration while LiIn(Cr$_{1-x}$Rh$_x$)$_4$O$_8$, which changes the electronic structure, appears to have almost no effect. This would suggest that the magnetic frustration is insensitive to the electronic properties and mostly dependent on the structure of the material.

%\section{Conclusions}

We have performed a systematic investigation of the effects of chemical substitution on the breathing pyrochlore LiIn(Cr$_{1-x}$Rh$_x$)$_4$O$_8$. From measurements of magnetic susceptibility and specific heat, we do not see any conclusive evidence of non-Fermi liquid behavior. However, from magnetization a broad feature centered at roughly 40 K is slowly suppressed with increasing Rh, until $x$ = 0.1 the feature is extremely broad and difficult to distinguish. From specific heat a peak at roughly 15 K for $x$ = 0 is slightly suppressed with initial Rh substitution, staying at 14 K for $x$ up to 0.1 but by $x$ = 0.2 the feature is completely suppressed. Furthermore, we find that the change in the electronic configuration or chemical pressure cannot fully explain response of $T_p$. However, from these measurements we find that using different chemical substitution can be used tune the amount of frustration which will be of great use for future attempts at uncovering new and enhanced magnetically frustrated systems.

%\bibliographystyle{cpl}
%\bibliography{cite1}

\end{document}